# Auger photoemission as a laser-like coherent cathode

Yushan Zeng[1], Bin Zhang[2,3], Kecheng Cao[3], Xiao-jing Liu[3], Yiming Pan[3]

1. State Key Laboratory of High Field Laser Physics and CAS Center for Excellence in Ultra-intense Laser Science, Shanghai Institute of Optics and Fine Mechanics, Chinese Academy of Sciences, Shanghai, China
2. Department of Electrical Engineering Physical Electronics, Tel Aviv University, Ramat Aviv 6997801, Israel
3. School of Physical Science and Technology and Center for Transformative Science, ShanghaiTech University, Shanghai 200031, China

**Abstract**

In pursuit of quantum advancements across disciplines, a bright and coherent electron source is expected to be a cornerstone of diverse applications including electron microscopy, laser accelerators, and free electron lasers. Current cathodes, such as cold field and photoemission, can generate high-quality electron beams with different cathode materials, geometric configurations, and laser excitation profiles, but their maintenance of both quantum coherence and high beam brightness suffers from the space-charge repulsion of many electrons. Here, we propose a new mechanism to provide collective emission of coherent electrons based on Auger photoemission. Our approach leverages a photon-induced four-level Auger process that necessitates a combination of photoemission and Auger recombination. The Auger electrons, energized through a recycling process of photoelectrons, emit collectively into the vacuum as secondary electrons. We compare coherent and incoherent Auger photoemission, identifying that the working condition of the coherent photoemission requires population inversion, akin to the four-level laser system. Our work provides insights for experimental realization and nanofabrication of Auger photocathodes, addressing a critical need in advancing quantum technologies relating to correlated coherent sources.

Coherence being intrinsic to the property of a wave (or quantum wave) provides unprecedented vistas of many micro/macroscopic matter properties and interactions. In 1960, the invention of the laser brought about coherent light sources, which then propelled breakthroughs in nonlinear optics, quantum optics, attosecond physics, and quantum information processing. To account for the remarkable coherence of laser light, in 1971, H. Haken et al's insight underscored the role of a non-equilibrium phase transition[1]. It is the cooperative emission from atomic dipoles that delineates coherent radiation of a laser from the radiation of a lamp[2-4]. Today, while lasers are still catalyzing further developments, the exploration of coherent waves has extended beyond photons to other matters - electrons, protons, and neutrons, mostly treated as particles. For our concern, analogous to the second-order phase transition of a superconductor, some intriguing questions remain to be answered: Can electrons be coherently emitted from a cathode in a way analogous to the coherent light from a laser? How to distinguish the coherence of such an electron source regarding their Fermionic character and Coulomb repulsion, and is there a cooperative mechanism at play?

The quantum nature of electrons, inherently as fermions, necessitates a distinct perspective on defining coherence in comparison with photons. Their difference recedes at the level of a single-particle source. However, while generating a single electron wavefunction with high spatial (or temporal) coherence and low emittance has been realized, the challenge aggravates with multi-electron sources. Being charged fermions, the coherence of an electron beam is predominantly constrained by Coulomb repulsion and Pauli's exclusion principle[5-8]. The longitudinal Coulomb repulsion can be suppressed strategically by propelling electrons to high-speed relativistic energy. Meanwhile, beam techniques by laser control[9-12] are being employed to push the beam emittance of electrons to the quantum limit in phase space, defined by the Heisenberg uncertainty. These approaches provide high-quality beam that can interfere with itself after passing different pathways[6,13,14], whose visibility, however, strengthens at the expense of subordinating the beam brightness. While Heisenberg-limited electron sources have been extensively studied, many phenomena that require phase-correlated electrons within a multi-particle beam are still not well understood. As such, the fundamental generation of correlated coherent electrons still awaits a new mechanism of cooperative emissions as the photons experience in a lasing regime.

In contrast to common cathodes that are commercially available, the recent advancements in exploring coherent electron sources have taken several approaches: attosecond strong-field photoemission[15,16], and resonant tunneling through discrete energy states inside a superconducting nanotip[17,18], quantum dots[19], and carbon nanotubes[20,21]. Both these sources rely on a nanostructured cathode, but their mechanism of electron emission differs: the field emission relies on strong-field ionization and scattering process, whereas the resonant tunneling emits almost monochromatic electrons through a quantum state that mediates the normal Fowler-Nordheim emission model[22-24]. Particularly, the recent exploration of new materials has discovered anomalous coherent electron emission from perovskite oxide[25] and possibly Cooper pairs of a superconducting niobium nanotip field emitter[18]. Looking ahead, the engagement of free-electron laser and radiations, electron microscopy and diffraction, and electron accelerators toward the quantum realm, governed by quantum mechanics, is approaching. In this pursuit of the quantum future, the quantum engineering of electron wavefunctions becomes crucial for further implementations, with an emphasis on bright and highly correlated coherent sources that are urgently required.

In this letter, motivated by the quest for a laser-like coherent electron source, we propose a semiclassical four-level model for photon-induced Auger electron emission. The Auger photoemission stands out from the common cathodes due to its intrinsic correlated emission dynamics via Coulomb interactions. The proposed Auger photoemission is similar to a four-level laser system, with the crucial difference of Auger recombination replacing radiative recombination. Our study identifies two steady-state photoemission regimes of a four-level system: (i) a weak light pumping intensity results in an ejected photocurrent proportional to the population of the pumped population, and (ii) in the saturation regime with sufficiently intense pumping, the emission current shows a quadratic reliance on the population – a phenomenon reminiscent of superradiance from a bunch of atoms[26]. The photoemission coherence arises from the population inversion and the non-radiative Auger process involving two electron collisions. By tuning the pumping light field, a transition from incoherent to coherent Auger photoemission is identified. Finally, we discuss the potential material realization as a new coherent Auger photocathode, which could supply future experiments that entail quantum electron sources.

**Four-level rate equations of Auger photoemission**. Regarding the coherence in the mechanism of electron emission induced by light, there are several pivotal factors[27,28]. For instance, in the photoelectric effect, the electron wavefunction inherits the optical phase from the pulsed laser, therefore each ejected electron carries the phase of the incoming photon when the photocathode is illuminated by a femtosecond light pulse[29,30]. The second factor is the pre-existence of correlated electrons in the cathode[7,8,31], which provides the opportunity for emitting correlated electrons by laser excitation or field emission while simultaneously preserving their coherence. In the photoelectric effect, as first explained by Einstein in 1905, electrons inside the material absorb one or multiple photons to overcome the material's work function. Their coherence degrades as electrons may be emitted at arbitrary optical phases.

Conversely, Auger recombination is characterized by an inherent correlated electron dynamic, wherein the electron transition happens through energy-conserving Coulomb interactions. Since the 'acceptor' electron receives energy from the recombination of another pair of electron and hole, the Auger electron's ejection is independent of the incident laser spectrum. Such features encourage an Auger photoemission theory which combines the Auger recombination and photoemission, shedding light on a substantial increase in the coherence of the photoelectrons compared to the existing cathode technologies.

To explore whether the nonradiative Auger process can indeed provide an essential edge in producing coherent electrons, we develop a four-level Auger photoemission system similar to the four-level laser system. Fig. 1a schematically depicts the energy level diagram of this proposed Auger electron emitter, with their energy states arranged as $E_3 > E_2 > E_1 > E_0$ and $N_3, N_2, N_1, N_0$ correspond to their respective populations. In analogy to the four levels of a laser, level 0 represents the ground state, while levels 1 and 3 are presumed to be short-lived ($\sim 10^{-7}s$), and level 2 is metastable ($\sim 10^{-3}s$) such as in Ti: sapphire and erbium-doped fiber amplifiers. The condition of population inversion is characterized as $N_2 > N_1$. When pumped by a light field, a large portion of the atoms (or electrons) is dispatched from the ground state to level 3, and then rapidly decays to the metastable level 2. The electrons on level 2 would gradually decay to level 1 through Auger recombination with half of the loss of the electron (emission, see Fig. 1c) rather than photon emission (Fig. 1b). Due to the short lifetime of level 1, the foregoing transition from level 2 to level 1 then leads to the population inversion between these two levels.

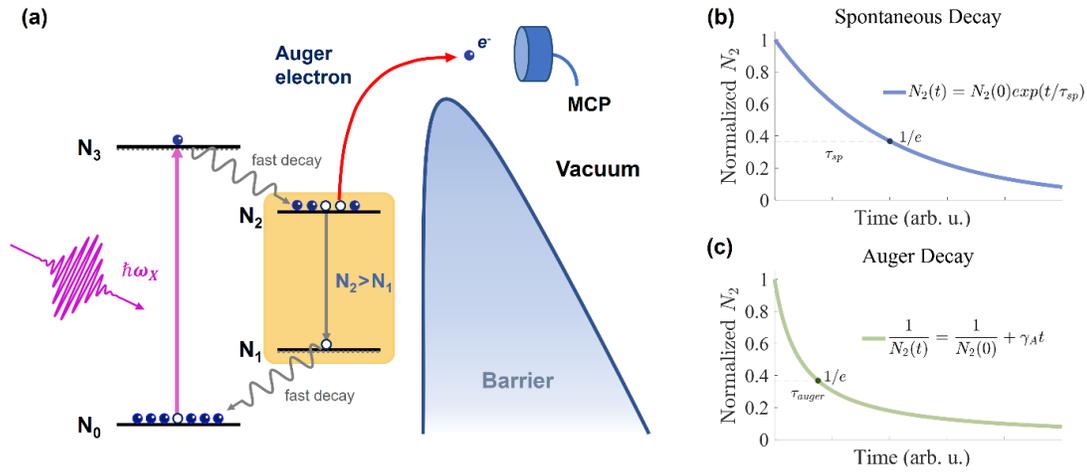

**Figure 1.** A schematic diagram of four-level coherent Auger photoemission. (a) Different from a typical four-level laser system, the photon recombination is replaced by an Auger recombination between level 1 and level 2. (b-c) Comparison of spontaneous relaxation processes between photon recombination ($\tau_{sp}$) and Auger recombination ($\gamma_A$). Note that the band-to-band transition should be forbidden with a negligible transition dipole moment, i.e., $|\langle 1|d \cdot E|2 \rangle| = 0$, so as to suppress the radiative recombination, and the high-order scattering process such as Auger recombination would be relevant. MCP, microchannel plate.

Similar to the laser system, the total populations of the Auger photoemission system predominantly occupy level 0 and level 2, as levels 1 and 3 remain nearly empty due to the short lifetimes. However, unlike the lasers, the key to the realization of Auger photoemission requires the absolute dominance of non-radiative transitions over radiative transitions between levels 2 and 1. Otherwise, the band-to-band recombination will transform the system into a traditional laser. Relevantly, this non-radiative process dominates in elements with a low atomic number and is universal in the semiconductors, offering chances of material realization of the coherent electron sources based on Auger photoemission. This part will be discussed later in the text.

To describe such an Auger photoionization system, a semiclassical approach is taken. Since our four-level Auger photoemission has comparable energy levels with the laser, we can easily derive the four-level rate equations of the populations in the Auger ionization:

$$\begin{aligned}\frac{dN_3}{dt} &= w_p(N_0 - N_3) - \frac{N_3}{\tau_3} \\ \frac{dN_2}{dt} &= \frac{N_3}{\tau_3} - \gamma_A N_2^2 \\ \frac{dN_1}{dt} &= \frac{\gamma_A N_2^2}{2} - \frac{N_1}{\tau_1} \\ \frac{dN_0}{dt} &= -w_p(N_0 - N_3) + \frac{N_1}{\tau_1}\end{aligned} \quad (1)$$

where $w_p$ is the stimulated emission coefficient depends on the pumping light intensity, $\tau_1, \tau_3$ are the lifetime of the levels 1 and 3, which are both short-lived (for example, $\tau_{1,3} \sim 10^{-7} s$). The relaxation of electrons on levels 1 and 3 is proportional to $N_1$ and $N_3$, similar to the rate equations of the laser. The Auger coefficient $\gamma_A$ characterize the collision between two electrons at metastable level 2. Crucially, for the three-particle process of Auger recombination described here, the Auger electron emission rate on level 2 is proportional to $N_2^2$ if the number of hole states on level 1 is constant. For equation (1.3), only half of the collision electrons drop to level 1 whereas the rest half of the population ($N_A$) has been scattered to other higher energy states as Auger electrons. Consequently, the combination of the four rate equations is not conservative, and the addition of the Auger electrons should be included to conserve the total population: $N = N_0 + N_1 + N_2 + N_3 + N_A$. The Auger electrons are emitted at the rate $\frac{dN_A}{dt} = \gamma_A N_2^2/2$.

In Auger recombination, the scattered electron obtains energy from the downward transition of a 'donor' electron (or multiple electrons in a high-order process) and is then kicked to a higher energy state. If the energy of the Auger electron is smaller than the work function of the material, the Auger electron will eventually relax back to the ground state by decaying through phonons and heating, such as in numerous semiconductors. Hence, to facilitate ionization of the Auger electron, a proper barrier is needed to allow the Auger electron to tunnel out of the material and finally be emitted as a free electron. To this end, the work function W should satisfy: $E_3 < W < 2E_2$, where $E_0$ is set as 0. Consequently, by incorporating all these considerations, the Auger photoemission process can be solved by many semiclassical tunneling processes, with an empirical formula of the Auger photocurrent:

$$J_{ph} = \beta \frac{dN_A}{dt} = \frac{\beta}{2} \gamma_A N_2^2 \quad (2)$$

where $\beta$ is the transmission coefficient of the tunneling process, which can be accounted by the Fowler-Nordheim tunnelling rate[32] (and other similar semiclassical models[33,34]) following a highly nonlinear dependence on the laser field intensity.

To solve the coupled rate equations, we adopt the steady-state approximation, i.e., $dN_i/dt \approx 0$. Hence, the populations on level 1 and level 3 can be derived as,

$$N_1 = \frac{w_p \tau_1}{1 + w_p \tau_3} N_0$$
$$N_3 = \frac{w_p \tau_3}{1 + w_p \tau_3} N_0 \tag{3}$$

Since the lifetimes $\tau_{1,3}$ are extremely short, the pumping light intensity $w_p$ and the lifetime of level 1 and 3 ($\tau_{1,3}$) satisfy: $w_p \tau_1, w_p \tau_3 \ll 1$, the populations on levels 1 and 3 are small enough to be neglected, i.e., $N_1 \simeq N_3 \simeq 0$. And $N_2, N_0 \gg N_A$. The conservation relation of the total population becomes,

$$N = N_0 + \frac{1}{1 + w_p \tau_3}(w_p \tau_1 + w_p \tau_3)N_0 + N_2 + N_A \simeq N_0 + N_2 \tag{4}$$

Under steady-state approximation, the population on level 2 can be solved by Eq. (1.2) and (4):

$$\frac{w_p}{1 + w_p \tau_3} N_0 - \gamma_A N_2^2 \simeq w_p(N - N_2) - \gamma_A N_2^2 = 0 \tag{5}$$

Hence, this yields the population of level 2,

$$N_2 = \sqrt{\left(\frac{w_p}{2\gamma_A}\right)^2 + \left(\frac{w_p N}{\gamma_A}\right)} - \frac{w_p}{2\gamma_A} = \frac{w_p}{2\gamma_A}\left[\sqrt{1 + \left(\frac{4N}{w_p/\gamma_A}\right)} - 1\right] \tag{6}$$

Equation (6) highlights the steady-state population of level 2 is always positive ($N_2 > 0$), ensuring that the population inversion between $N_2$ and $N_1$ holds.

To gain further physical insight with this formalism, two limits of pumping conditions are considered. We first deal with the scenario of a weak pumping field: $w_p/\gamma_A \ll N$, such that $N_2 = \sqrt{\left(\frac{w_p}{2\gamma_A}\right)^2 + \left(\frac{w_p N}{\gamma_A}\right)} - \frac{w_p}{2\gamma_A} \approx \sqrt{\frac{w_p N}{\gamma_A}}$. The Auger photoemission current is given by,

$$J_{ph} = \beta \frac{dN_A}{dt} = \frac{\beta}{2} w_p N, \tag{7}$$

This formula shows that the Auger photocurrent is proportional to the total population $N$, and the light intensity ($w_p$). It is interesting to note that the Auger photoemission is independent of the Auger coefficient $\gamma_A$, suggesting that in this weak pumping regime, the photoemission current is induced by the light pumping process instead of the Auger process.

Conversely, by increasing the pumping light intensity into the strong field regime: $w_p/\gamma_A \gg N$, we obtain $N_2 = \sqrt{\left(\frac{w_p}{2\gamma_A}\right)^2 + \left(\frac{w_p N}{\gamma_A}\right)} - \frac{w_p}{2\gamma_A} \approx N$. This expression makes explicit that almost the whole population contained within the system has been dispatched into the metastable level 2. In this case of the pump saturation, the Auger photoemission current is given by,

$$J_{ph} = \beta \frac{dN_A}{dt} = \frac{\beta \gamma_A N^2}{2} \qquad (8)$$

Equation (8) states that the Auger photoemission current, under the pumping of an intense light field, becomes independent of the pumping light intensity. This is what we refer to as the saturation regime. According to equation (8), another noteworthy feature is the photocurrent's quadratic reliance on the total population $N$, indicating a brand-new photoemission mechanism which is a hallmark of superradiance in coherent light emission. To be clear, we assert that the coherent Auger photoemission (Eq. 8) is the main achievement of the present work.

**Direct band-to-band transition versus Auger transition.** To distinguish the nonradiative Auger recombination and the radiative spontaneous transition, we make comparisons of their decay rates of the $N_2$. The rate equation from spontaneous recombination is $dN_2/dt = -N_2/\tau_{sp}$, while for saturated Auger recombination, the rate equation is $dN_2/dt = -\gamma_A N_2^2$. Correspondingly, we can obtain their respective population as a function of time as: $N_2(t) = N_2 e^{-t/\tau_{sp}}$, $N_2(t) = \left(\frac{1}{N_z} + \gamma_A t\right)^{-1}$, where $N_z = N_2(0)$ is the initial population of the level 2. Fig. 1b and 1c compare the population rate's difference between photon recombination and Auger recombination. In many cases, the radiative photon transition is more direct and effective compared to the Auger process. The photon recombination can be understood as a scattering process between an electron at level 2 and one hole at level 1, whereas the Auger recombination process is a high-order scattering process that requires at least one collision or scattering between two electrons at level 2 (or around level 2) and one hole at level 1.

An intuitive way to diminish the photon recombination and enhance the Auger recombination is to construct two levels that cannot allow a direct band-to-band transition, Otherwise, the band-to-band recombination will dominate the system. This leads to the requirements for the operating levels: $|\langle 1|\hat{d} \cdot E(t)|2\rangle|^2 = 0$, where $\hat{d} \cdot E$ is the dipole transition operator corresponding to the interaction between an electron and a radiation field. The scattering matrix is zero if the wavefunctions of level 1 and level 2 are simultaneously odd or even. This provides the necessary clue for searching for the material realization of the laser-like coherent Auger electron sources.

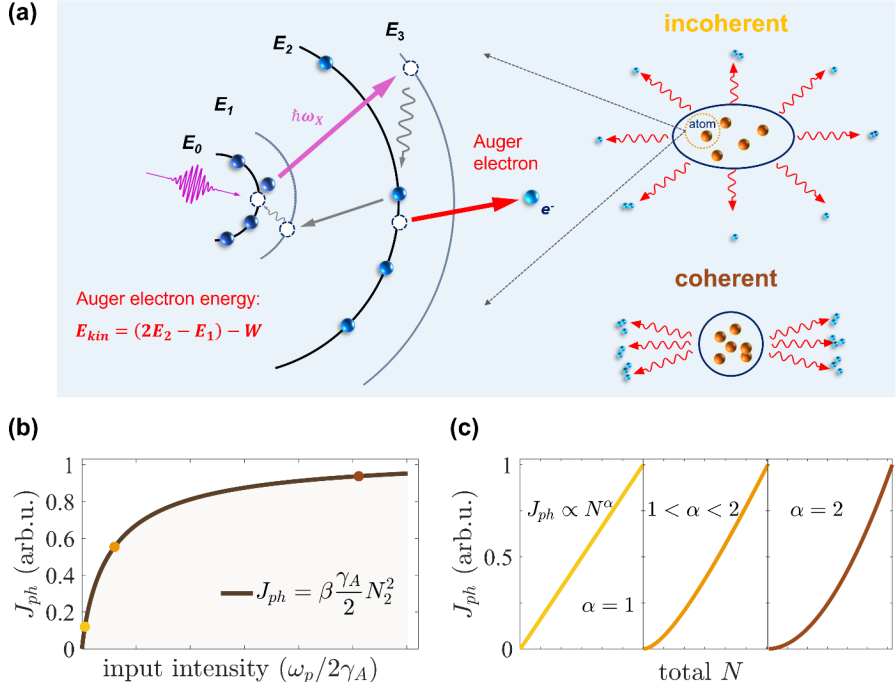

**Figure 2.** The Auger transition from incoherent to coherent photoemissions. (a) The transition shows the similarity with the superradiant transition of atoms from incoherent to coherent radiation of light. (b) Photoemission current as a function of the input light intensity. The colored dots represent different situations corresponding in (c), respectively. (c) We compare three regimes of the Auger photoemission: weak pumping field regime (yellow line); mediated pumping field regime (orange line); and saturated pumping field regime (brown line). The scaling with the population is linear in the weak pumping regime, and quadratic in the saturated pumping regime.

**Origin of coherence in Auger photoemission.** We introduce a collective electron source mechanism to interpret the coherent and incoherent Auger photoemission within a population of $N$ atoms. Consider that each atom can be individually ionized and the total Auger electron state is described by $\psi(t) = \sum_{i=1}^{N} \psi(t - t_i)$, where $t_i$ represents the ionization time of the i$^{\text{th}}$ atom. From this, we derive the spectral density distribution of the total Auger electron: $\rho_{ph} = |\psi(\omega)|^2 \sum_{i,j=1}^{N} e^{i\omega(t_i - t_j)}$. Averaging of this yields the photoemission current:

$$J_{ph} \propto \langle \rho_{ph} \rangle = |\psi(\omega)|^2 \sum_{i,j=1}^{N} \int dt_i dt_j \, f(t_k) f(t_m) e^{i\omega(t_i - t_j)} \qquad (9)$$
$$= |\psi(\omega)|^2 (N + N(N-1)|f(\omega)|^2).$$

Here, $f(\omega) = \int dt_i f(t_i) e^{i\omega t_i}$ represents the Fourier transform of the ionized electrons' distribution. The first term corresponds to incoherent photoemission, proportional to the population ($N$) of the metastable level 2, whereas the second term represents the coherent Auger photoemission that scales quadratically with $N$ (Fig. 2c, brown line). The function $|f(\omega)|^2$ characterizes the coherence of the Auger photoemission, in analogy with Dicke superradiance[26], as illustrated in Fig. 2a (coherent situation). The underlying coherence of Auger electrons at the saturated regime emerges from the energy-conserving collision between collective electrons. This is further augmented by the photoemission recycling process, which harvests the recombination energy to eject a secondary electron at the upper level not directly excited by the incident light (Fig. 2a). As a result, the Auger photoemission current remains coherent in the saturated regime, despite the collective emission from metastable level 2 in both regimes.

**Material realization of coherent electron sources.** While Auger interactions are prevalent in the dynamics of the inner shell and find applications in Auger electron spectroscopy, radiation therapy[35], and light-emitting systems[36,37], the exploration of their Coulomb origin for coherent sources remained largely unexplored. A major barrier to be tackled is the efficiency of Auger processes compared to radiative transitions and other intraband relaxation mechanisms, such as phonon-assisted carrier cooling[38,39]. Envisioning the coherent Auger photoemission as a four-level laser system, we explore potential material realizations, including atomic systems and semiconductors with discrete energy bands.

The deep electron orbitals close to the nucleus have a strong electron-electron interaction, which facilitates a dominant Auger process, particularly in elements with low atomic numbers. Notably, noble gases like Neon and Krypton have been determined with an ultrafast Auger decay rate of several femtoseconds[40,41], featuring an extremely large Auger probability over other processes. Moreover, the ionized electrons are expected to possess higher initial kinetic energy than those emitted from conventional cathodes, owing to their stronger attraction towards the nucleus. However, the requirement for high-energy photons in the extreme ultraviolet (XUV) range poses a challenge to the widespread adoption of Auger photoemission sources.

We therefore consider semiconductors as a promising alternative for creating a reservoir of materials suitable for laser-like Auger photoemission. In semiconductors, phenomena such as radiative recombination and intraband cooling often outpace the Auger effect, impeded by the screened Coulomb interaction and the limitations of energy and momentum conservation[42]. Indeed, in numerous semiconductors, the Auger process is deemed deleterious, with efforts directed toward its mitigation to improve the light-emitting efficiency in applications like solar cells and diodes[43,44]. Yet strategies to boost Auger efficiency remain insufficiently explored. Beyond the constraints imposed by selection rules on the transition matrix, potential methods for augmenting Auger efficiency in semiconductors include doping[45], bandgap engineering[46,47], and confining carrier space, within a quantum dot[48,49]. Additionally, in addition to electron ejection through barrier tunneling, the 'acceptor' electron can also accumulate energy from multiple electron-hole combination events. In this respect, several studies utilizing colloidal quantum dots have showcased spin-exchange-enhanced Auger ionization via doping with a magnetic element [50,51].

We note that, a recent observation revealed an unexpected photoemission peak intensity in the single crystals of the perovskite oxide SrTiO3 (100)[25,52] at low temperatures, presenting discrete photoemission spectra. This emergence of coherence, attributed to the secondary emission of these failed photoelectrons alongside the conventional three-step photoemission theory, resonates with a similar combination of the Auger process and photoemission outlined in our work. These significant experimental and material advances, together with our comprehensive framework for coherent Auger photoemission, underscore a promising yet largely unexplored domain within the field of photocathode quantum materials and coherent electron sources.

**Conclusion.** In short, we propose a novel mechanism for generating coherent Auger photoemission. The proposed photoemission process is analogous to the configuration of a four-level laser but utilizes nonradiative energy-conserving Coulomb interactions. This cooperative ionization enables a surge in the emitted electrons that is reminiscent of Dicke's superradiance. Notably, in a semiconductor laser diode, incoherent currents are injected, resulting in a coherent laser field. Conversely, our Auger photoelectron emitter enables the incoherent light pumping of coherent electrons. Drawing advantages of the Auger photoemission framework, we anticipate that our work will shed light on developing new photocathode electron sources for future technologies and applications.

**Acknowledgement**

We thank the discussions with Peter Hommelhoff, Michael Kruger for their insightful discussions. Y.Z. acknowledges the support of the National Natural Science Foundation of China (NSFC) (grant No. 12104471). Y.P. acknowledges the support of the NSFC (Grant No. 2023X0201-417-03).